\documentclass{article}

\usepackage{arxiv}

\usepackage[utf8]{inputenc} 
\usepackage[T1]{fontenc}    
\usepackage{hyperref}       
\usepackage{url}            
\usepackage{booktabs}       
\usepackage{amsfonts}       
\usepackage{nicefrac}       
\usepackage{microtype}      
\usepackage{lipsum}         
\usepackage{graphicx}
\usepackage{natbib}
\usepackage{doi}

\usepackage{lipsum}
\usepackage{appendix}
\usepackage{tabularx}
\usepackage{color}
\usepackage{ltxtable}
\usepackage{amsmath}
\usepackage{cleveref} 
\usepackage{amssymb}
\usepackage{utfsym}
\usepackage{tablefootnote}
\usepackage{float}

\title{COLA: A Scalable Multi-Agent Framework For Windows UI Task Automation}

\date{November 2, 2024}

\author{
  \textbf{Di Zhao\textsuperscript{1}},
  \textbf{Longhui Ma\textsuperscript{1}},
  \textbf{Siwei Wang\textsuperscript{2}},
  \textbf{Miao Wang\textsuperscript{2}},
  \textbf{Zhao Lv\textsuperscript{2}}
  \\
  \textsuperscript{1}National University of Defense Technology, 
  \textsuperscript{2}Academy of Military Sciences
  \\
  \small{
      {zhaodi@nudt.edu.cn}
  }
}

\begin{document}
\maketitle

\begin{abstract}
With the rapid advancements in Large Language Models (LLMs), an increasing number of studies have leveraged LLMs as the cognitive core of agents to address complex task decision-making challenges. Specially, recent research has demonstrated the potential of LLM-based agents on automating Windows GUI operations. 
However, existing methodologies exhibit two critical challenges:
(1) static agent architectures fail to dynamically adapt to the heterogeneous requirements of OS-level tasks, leading to inadequate scenario generalization;(2) the agent workflows lack fault tolerance mechanism,  necessitating complete process re-execution for UI agent decision error.  
To address these limitations, we introduce \textit{COLA}, a collaborative multi-agent framework for automating Windows UI operations.
In this framework, a scenario-aware agent Task Scheduler decomposes task requirements into atomic capability units, dynamically selects the optimal agent from a decision agent pool, effectively responds to the capability requirements of diverse scenarios.
The decision agent pool supports plug-and-play expansion for enhanced flexibility.
In addition, we design a memory unit equipped to all agents for their self-evolution.
Furthermore, we develop an interactive backtracking mechanism that enables human to intervene to trigger state rollbacks for non-destructive process repair.
Our experimental results on the GAIA benchmark demonstrates that the \textit{COLA} framework achieves state-of-the-art performance with an average score of 31.89\%, significantly outperforming baseline approaches without web API integration. Ablation studies further validate the individual contributions of our dynamic scheduling. The code is available at \url{https://github.com/Alokia/COLA-demo}.

\end{abstract}

\section{Introduction}

In recent years, technologies based on Large Language Models (LLMs) have advanced rapidly, demonstrating significant potential in language dialogue and problem-solving \citep{wang2024survey, guo2024large}. 
More complex multi-modal models (MLLMs), such as GPT-4v \citep{achiam2023gpt}, GPT-4o, and Gemini \citep{team2023gemini}, introduce a visual dimension, expanding the capabilities of LLMs and demonstrating outstanding capabilities across a broader range of fields \citep{guo2024large}. 
High-capacity LLMs, MLLMs often serve as the main backbone of agents tested in specialized fields, such as software development \citep{hong2023metagpt, chan2023chateval, li2023camel}, social simulation \citep{park2023generative, gao2023s}, and gaming \citep{akata2023playing, wang2023voyager, tan2024towards}.

As a practical multi-modal application scenario, automated tasks on personal computers are emerging as a key area of technological advancement in AI system assistants \citep{niu2024screenagent, wu2024copilot, zhang2024ufo}. 
Users interact with computers and access information primarily through the User Interface (UI) or Graphical User Interface (GUI) of software applications. 
However, due to limited screen recognition, operation, and location capabilities, existing MLLMs face challenges in this scenario \citep{zhang2024large, wang2024gui}. 
To address this, existing work leverages MLLM-based agent architecture to endow MLLMs with various capabilities for perceiving and operating computer device \citep{song2024mmac, wang2024sibyl, nguyen2024dynasaur, hfagent2024}. 
UFO \citep{zhang2024ufo} introduces a dual-agent system, utilizing an AppAgent to manage application operations and decision-making across various scenarios.
Nevertheless, this approach struggles to handle more complex GAIA datasets \citep{mialon2024gaia}.
MMAC \citep{song2024mmac} develops agents for four distinct tasks: programming, screen semantic recognition, video analysis, and general knowledge. 
However, the system design suffers from limited scalability and lacks flexibility. 
Any error in the execution process necessitates a complete restart, which can significantly hinder efficiency and adaptability in practical applications.

Computer task scenarios are inherently complex, involving a range of specialized skills such as coding, information retrieval, file management, and system configuration, often necessitating the integration of information across multiple applications.
Current research \citep{zhang2024ufo, song2024mmac, wu2024copilot} typically utilizes orchestrated static agent systems to manage these tasks, but such approaches are inefficient when applied to more complex scenarios. 
This limitation underscores a significant gap in adaptability and scalability, which are essential for addressing increasingly sophisticated computer task environments.

\begin{figure} 
\centering
\includegraphics[width=\textwidth]{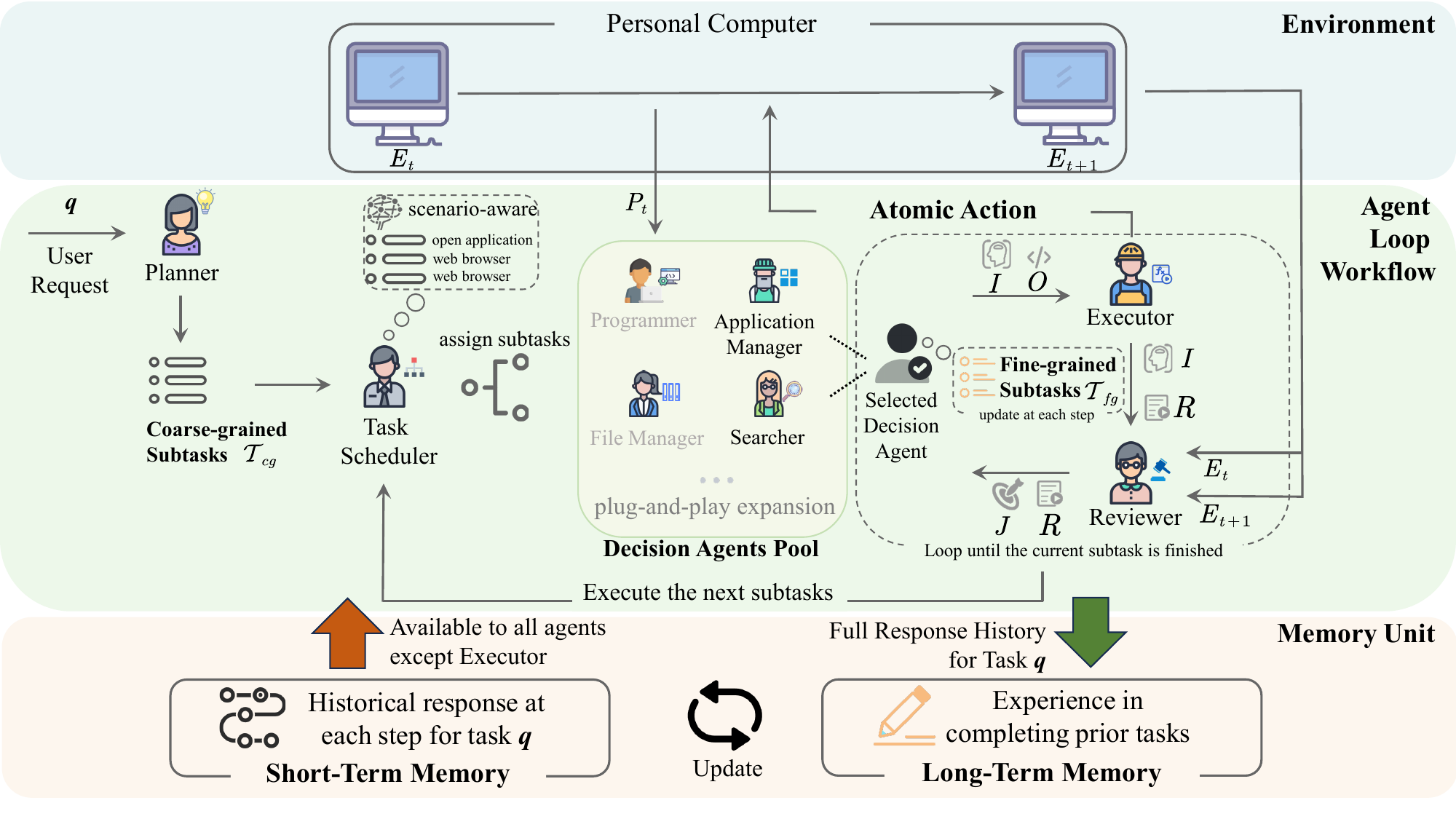}
\caption{\textbf{An illustration of the \textit{COLA} multi-agent framework}. In the first step, Planner takes request $q$ from user and decomposes it into a sequence of coarse-grained subtasks ($\mathcal{T}_{cg}$). Task Scheduler then dynamically selects optimal decision agents through scenario-aware matching. Selected Decision Agents subsequently perform hierarchical task refinement, utilizing their domain-specific expertise to decompose assigned subtasks into fine-grained subtasks ($\mathcal{T}_{fg}$), giving an atomic action $O$ and an intention $I$ to execute that action. Executor executes it and obtains the environmental feedback result $R$. Finally, the Reviewer evaluates the success of the action based on the environment $E_t$, $E_{t+1}$ before and after execution, the intention $I$ and the result $R$. The judgment $J$ is then sent back to the selected Decision Agent. This cyclic refinement continues until all subtask requirements are satisfied, with the Task Scheduler orchestrating inter-subtask transitions. Throughout the process, humans can intervene in the workflow at any time, providing guidance to correct the agent's response.}
\label{fig:frame}
\end{figure}

In this work, we introduce \textit{COLA}, a scalable and flexible multi-agent framework for Windows operating system assistants.
\textit{COLA} incorporates five specialized agent roles: \textbf{planner}, \textbf{task scheduler}, \textbf{decision agent pool}, \textbf{executor}, and \textbf{reviewer}.
The decision agent pool consists of agents with domain-specific expertise, each tailored to handle specific tasks such as web browsing, file manipulation, programming, and others.
We utilize the task scheduler to select the most appropriate decision agent from the pool when faced with different scenarios.
The decision agent pool supports plug-and-play extensions, allowing users to design specialized agents for a particular scenario, thus eliminating the need to rewrite the framework and completing the extension of capabilities.
Furthermore, to avoid the framework requiring to execute from scratch when agent responds anomalously, inspired by Swarm \footnote{\url{https://github.com/openai/swarm}}, we develop an interactive backtracking mechanism.
This mechanism allows users to revert to any previous response state of an agent, provide corrective guidance, and resume workflow execution from that state.
The entire process is illustrated in Figure~\ref{fig:frame}. 

Our summarized contributions are as follows:
\begin{itemize}
\item 
We propose \textit{COLA}, a scalable and flexible collaborative multi-agent framework that implements hierarchical task resolution through five specialized agent roles for complex computer task automation. 
Each agent is equipped with two types of memory unit: a short-term memory unit that updates as tasks progress, aiding the agent in understanding task developments, and a long-term memory unit that logs completed work, continuously enhancing the agent's capabilities.

\item 
We formalize the decision agents as a pool of various agents with domain-specific expertise, and use a special Task Scheduler to perceive the task scenario and dynamically select the optimal decision agent from the pool.
Furthermore, we develop an interactive backtracking mechanism that allows human users to intervene to trigger state rollbacks for non-destructive process repair.
\item 
Experimental results on the GAIA benchmark demonstrate state-of-the-art performance, validating \textit{COLA}'s effectiveness in handling complex computer task automation scenarios.
\end{itemize}

\section{Related Work}

\subsection{LLM based Agents}

In recent years, LLM-based agents have been considered as a promising approach to achieving artificial general intelligence (AGI) \citep{wang2024survey}. 
It significantly expands the capabilities of LLMs, empowering them to engage in planning, memorization and executing actions \citep{guo2024large}.
This enhancement allows LLMs to accomplish more complex tasks by mimicking human-like thinking processes and the ability to interact with the environment. 
LLM-based agents have been designed and applied in a variety of domains, including social simulation \citep{park2023generative, gao2023s}, gaming \citep{akata2023playing, tan2024towards, wang2023voyager}, code generation \citep{hong2023metagpt, chan2023chateval}, etc.

Inspired by human-team collaboration, multi-agent systems are receiving increasing attention. 
\citep{qian-etal-2024-chatdev} presents an end-to-end framework for software development that utilizes multiple agents to collaborate on software development tasks. 
\citep{tao2024chain, chan2023chateval, subramaniam2024debategpt} explores the potential of enhancing the quality of generated content through the use of multiple agents participating in debates. 
\citep{hong2023metagpt} presents a groundbreaking framework for encoding Standardized Operating Procedures (SOPs) into prompt sequences to enhance collaboration.

\subsection{LLM-based UI Operation Agent}

The utilization of LLM-based agent systems for navigating and controlling graphical user interfaces (GUIs) has emerged as a novel and rapidly expanding research area. 
\citep{yan2023gpt} proposes a multi-modal agent based on GPT-4V for navigating mobile applications by directly inputting the screenshot. 
Mobile Agent \citep{wang2024mobile-a} integrates Optical Character Recognition (OCR) technology to enhance the agent's visual understanding. 
Furthermore, Mobile Agent v2 \citep{wang2024mobile-b} improves single-agent to multi-agent for better performance on multiple tasks. 
It defines three types of agents: Planning Agent, Decision Agent and Reflection Agent, and all actions are given by a Decision Agent. 
UFO \citep{zhang2024ufo} utilizes the Python package pywinauto to inspect the UI controls and implement actions. It defines two types of agents: HostAgent and AppAgent, where all decisions are made by the AppAgent. 
However, these approaches rely on a single agent to make decisions for all tasks, which limits their scalability. 
For complex tasks, it is challenging for one agent to handle decisions across all scenarios.
To address this limitation, in this paper, we propose the \textit{COLA} framework, which treats Decision Agent as a scalable pool of specialized agents, each designed to handle specific tasks.
The framework assigns the tasks to the most appropriate agent, enabling decision-making based on the current scenario.

\section{The COLA Framework}

In this section, we will provide a detailed overview of the \textit{COLA} architecture.
The operation of \textit{COLA} is sequential and iterative, and its process is depicted in Figure~\ref{fig:frame}.
We design the memory unit to enhance the agent's comprehension of task progress and its capacity to self-evolution based on prior experience.
Additionally, we develop an interactive backtracking mechanism that enables non-destructive process repair and avoids workflow execution from scratch.
The prompts for each agent are described in the Appendix~\ref{app:prompts}.

\subsection{Visual Perception and Interaction}
\label{ss:visual_perception}

\begin{figure} 
\centering
\includegraphics[width=\textwidth]{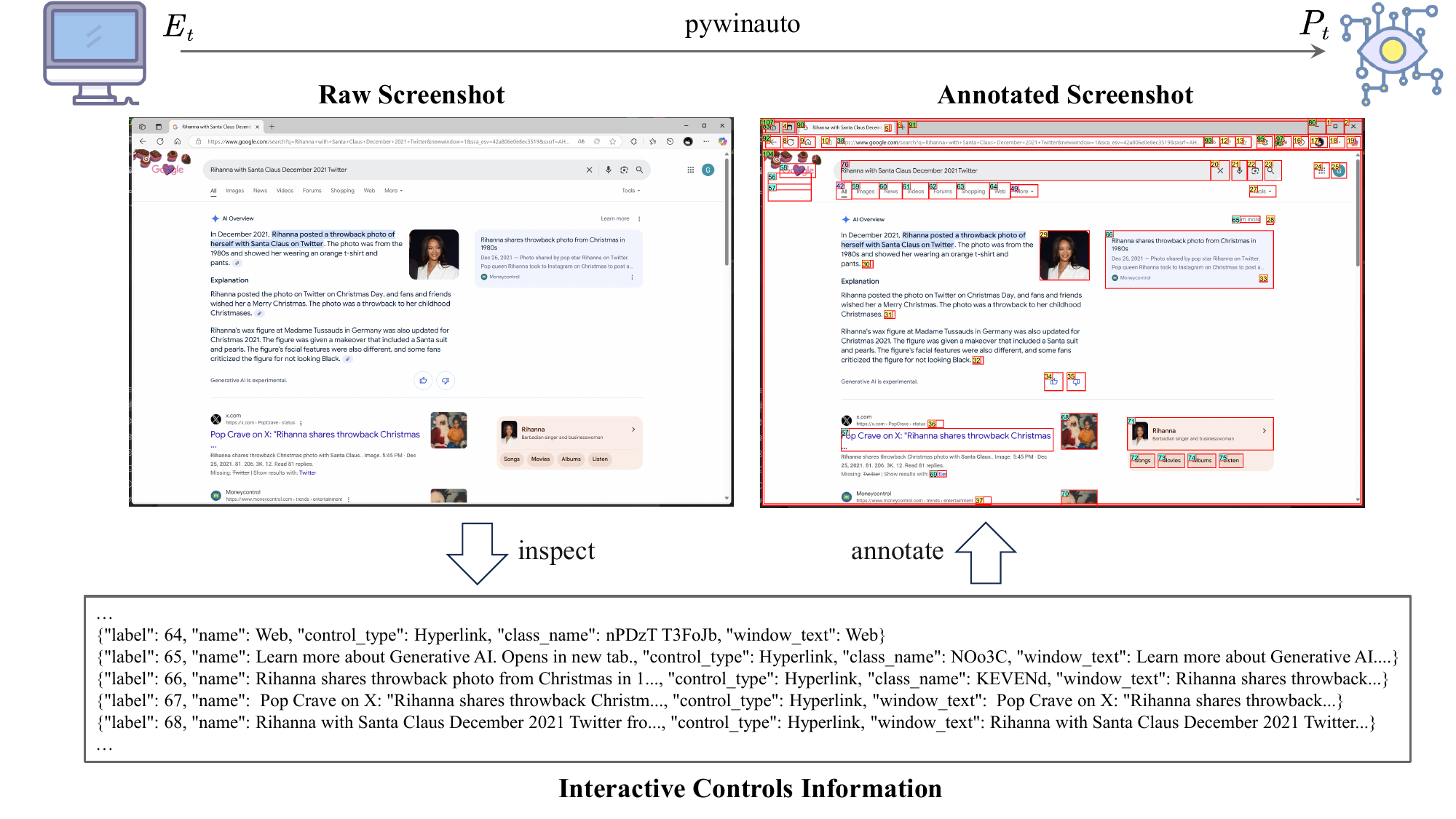}
\caption{A visual perception example for Microsoft Edge with information provided by \textit{pywinauto}. The raw screenshot, annotated screenshot and interactive controls information make up the visual perception component $P_t$.}
\label{fig:vp}
\end{figure}

Screen recognition remains challenging even for state-of-the art MLLMs. 
Making accurate decisions from MLLMs based solely on screenshots is particularly difficult. 
Therefore, we adopt the same methodology employed by UFO \citep{zhang2024ufo}, utilizing the Python package \textit{pywinauto} \citep{bim2014application} to inspect interactive controls within applications. 
We define the process of visual perception as $\mathcal{F}$, which is formally represented by the following equation:
\begin{equation}
P_t = \mathcal{F}(E_t)
\end{equation} 
where $E_t$ represents the screen state at step $t$. 
$P_t$ denotes the visual perception component as illustrated in Figure~\ref{fig:vp}. 

In subsequent developments, agents requiring screen information for decision-making will incorporate the visual perception component as part of their prompt.
This enhancement significantly improve their perception and comprehension of the current desktop environment, fostering more accurate and effective decisions across various applications.

To interact with the computer environment, we developed eight actions, as detailed in Appendix~\ref{app:actions}.
We design a domain mechanism for each action so that only agents registered in the domain can use the action. 
This design paradigm effectively manages the agent's capabilities while minimizing the complexity associated with expanding the action space.
By employing these predefined actions, the agent is able to interact with the computer system. 
Users can custom actions to meet their specific requirements and configure the domain in which agent can recognize and apply them.

\subsection{Memory Unit For Self-Evolution}

Due to the fact that the computer operation task involves numerous scenarios and requires multi-step sequential operations, the LLM-based agent needs to have the capacity to learn from past experiences and be able to articulate progress on current tasks. 
Inspired by how humans become increasing effective and efficient in operating computer, we maintain a long-term memory and a short-term memory.

\paragraph{Long-Term Memory}
The long-term memory (denoted as $LT$) preserves a complete record of prior task executions, facilitating the agent's ability to learn from its past experiences.
To enable the agent to access records in $LT$, we introduce a retrieval function $\mathcal{L}$:  $\mathcal{Q}\times \mathbb{N}\rightarrow 2^{LT}$, where $\mathcal{Q}$ represents the space of queries and $\mathbb{N}$ denotes the set of positive integers. 
For each record, a summary is generated. 
These summaries are subsequently embedded to create a set of indices corresponding to the records.
Given a query $q\in \mathcal{Q}$ and an integer $n\in \mathbb{N}$, the function $\mathcal{L}(q, n)$ embeds the query using the same embedding applied to the summaries, then computes the cosine similarity between the query’s embedding and the embedding of each record's summary. 
The top-$n$ records in $LT$ with the highest similarity scores are returned to the agent as part of its prompt. 
For convenience, we denote the top-$n$ records associated with query $q$ at step $t$ as $LT^n_t$.

\paragraph{Short-Term Memory}
The short-term memory (denoted as $ST$) retains the historical responses generated at each step of the current task, forming a sequence of operations: $ST_t=\left\{ st_1,\ st_2,\ ...,\ st_t \right\}$, where $st_t$ represents the response produced by the agent at step $t$. 
However, including the entire $ST_t$ in the prompt may lead to increased computational costs.
To mitigate this, only the most recent $m$ responses are utilized: $ST^m_t=\left\{ st_{t-m+1},\ st_{t-m+2},\ ...,\ st_t \right\}$.
Each agent is equipped with both types of memory, which are not shared among agents.

Each agent possesses an independent memory storage space.
The short-term memory records responses at each step of task execution, facilitating the agent's understanding of task progress.
Upon task completion, decisions are stored in long-term memory, enabling the agent to recall past decision-making processes when encountering similar tasks in the future, thereby expanding its strategic perspective.
When a query is received, the relevant long-term memory ($LT^n_t$) and short-term memory ($ST^m_t$) are integrated into the prompt to improve the agent's decision-making capacity.

\subsection{Hierarchical Multi-Agent Framework}

Numerous studies have demonstrated that collaboration among multiple agents possessing diverse skills can enhance task performance \citep{hong2023metagpt, chan2023chateval, wang2024mobile-b, wu2023autogen}. 
In \textit{COLA} framework, we established five types of agents: Planner, Task Scheduler, Decision Agent Pool, Executor, and Reviewer. 
The Decision Agent Pool, in particular, comprise a series of scalable agents, each with specialized skills, including the Application Manager, File Manager, Programmer, and Searcher, as depicted in Figure~\ref{fig:frame}.
The inputs and outputs of each agent are detailed as follows.

\subsubsection{Planner}

The planner plays a crucial role in managing the workflow process by breaking down user requests into subtasks, establishing an organized and methodical foundation for task execution. 
The planner initially generates coarse-grained subtasks, which are then further refined into fine-grained subtasks by subsequent decision agent.
This hierarchical planning approach is particularly effective for handling complex and variable tasks.

We define the coarse-grained subtasks generated by planner as $\mathcal{T}_{cg}$. Given a user request $q$, this process is represented by the following formula:
\begin{equation}
\mathcal{T}_{cg} = \left\{ s_1,\ s_2,\ ...,\ s_k \right\} =PL\left( q, \ LT^n_t,\ ST^m_t \right)
\end{equation}
where $PL$ represents the LLM of planner, each $s_k\in \mathcal{T}_{cg}$ represents a coarse-grained subtask. 

The planner's high degree of functional decoupling from other agents affords significant flexibility, enabling it to effectively employ a variety of reasoning strategies, such as COT \citep{wei2022chain}, TOT \citep{yao2024tree}, or multi-agent debate \citep{liang2023encouraging}, to improve response performance.

\subsubsection{Task Scheduler}

The task scheduler is designed to identify the capabilities required for each coarse-grained subtask generated by the planner.
It then assigns these subtasks to the appropriate agents based on their specialized descriptions in the decision agent pool.
This process is represented by the following formula:
\begin{equation}
\begin{split}
    \mathcal{D} &=\left\{ \left( role_1,\ rt_1 \right) ,\ \left( role_2,\ rt_2 \right) ,\ ...,\ \left( role_k,\ rt_k \right) \right\}\\
    &=TS\left( \mathcal{T}_{cg},\ DA_{desc},\ LT^n_t,\ ST^m_t \right) 
\end{split}
\end{equation}
where $TS$ represents the LLM of task scheduler, $role_k$ refers to an agent in the decision agent pool, $rt_k$ denotes the coarse-grained subtasks assigned to $role_k$, and $DA_{desc}$ represents the description of the specialties of all agents in the decision agent pool.

After the assignment $\mathcal{D}$ is generated, each agent $role_k$ is sequentially tasked with performing its assigned subtask $rt_k$.

\subsubsection{Decision Agent Pool}

Previous approaches relied on single or multiple fixed agents to make decisions in specific scenarios \citep{zhang2024ufo, song2024mmac, wang2024mobile-b}, but this static agent architecture fails to dynamically adapt to the heterogeneous demands of operating system tasks, struggling to manage the complexity and variety of computer tasks.
In contrast, drawing inspiration from the Mixture of Experts (MoE) model \citep{jacobs1991adaptive}, we formalize the decision agent as a scalable pool comprising agents with specialized capabilities, each tailored to distinct scenarios.
Each agent's expertise is represented by a natural language description, denoted as $DA_{desc}$.
When the task scheduler assigns subtasks, the selected agent $role_k$ from the pool completes the assigned subtasks $rt_k$ sequentially, based on the visual perception component $P_t$ discussed in Section~\ref{ss:visual_perception}.
This process is represented by the following formula:
\begin{equation}
\left( I, O, \mathcal{T}_{fg} \right) =DA_{role_k}\left( q, rt_k, P_t, J, LT^n_t, ST^m_t \right) 
\end{equation}
where the $DA_{role_k}$ represents the selected agent in decision agent pool; $O$ represents the action to be performed, and $I$ is the intention to perform the action; $\mathcal{T}_{fg}$ is a fine-grained list of subtasks that is regenerated for each execution to adjust the planning immediately; $J$ is the judgment given by reviewer.

This design paradigm, which dynamically assigns tasks through the task scheduler, enables plug-and-play scaling of the decision agent pool.
It allows users to customize both specialized agents and their associated actions using the domain mechanism described in Section~\ref{ss:visual_perception}.

Figure~\ref{fig:gaia} illustrates the skill requirements of the GAIA benchmark. In this study, we implemented four ad-hoc agents to meet these requirements, each described in natural language as follows:
\begin{itemize}
\setlength{\itemsep}{0pt}
\setlength{\parsep}{0pt}
\setlength{\parskip}{0pt}
    \item \textbf{Application Manager}: Can open applications such as browsers, explorers, chat software, etc.
    \item \textbf{File Manager}: Can open, create, and delete files, such as txt, xlsx, pdf, png, mp4 and other documents.
    \item \textbf{Searcher}: Can use an opened browser to search for information, open web pages, etc. Can also do everything related to web pages, such as playing videos in web pages, opening files, reading documents in web pages, and so on.
    \item \textbf{Programmer}: Possesses logical reasoning and analytical skills. Can reason to arrive at an answer to a question or write Python code to get the result.
\end{itemize}

\subsubsection{Executor}

The executor is responsible for directly interacting with the computer environment. 
Since its role is limited to executing actions, there is no need for a memory unit. 
This process is represented by the following formula:
\begin{equation}
    \left( E_{t+1},\ R \right) =Exec\left( O,\ E_t \right) 
\end{equation}
where the $Exec$ denotes the function by which the executor performs the action, $E_{t+1}$ represents the environment state after executing action $O$, and $R$ is the result of the action, which may be null, as in the case of actions like a mouse click.

Due to the potential dangers of operations within the computer environment, such as file deletion, we have decoupled the direct interaction functionalities of decision agents from the environment into a separate executor component. 
This decoupling facilitates subsequent research on restricting sensitive operations without necessitating modifications to other system components.

\subsubsection{Reviewer}

Due to the hallucination problem associated with LLMs \citep{liu2023aligning, gunjal2024detecting, cui2023holistic}, agents may generate unintended actions in certain scenarios. 
To address this, we design the reviewer to assess the validity of an action based on changes in the operating environment before ($E_t$) and after ($E_{t+1}$) the action is performed, as well as the intent ($I$) behind the action, generated by the decision agent. 
This process is formalized as follows:
\begin{equation}
    J=\text{Re}\left( E_t,\ E_{t+1},\ I,\ O, \ R \right) 
\end{equation}
where $Re$ represents the LLM of reviewer.

The reviewer evaluates the actions of decision agents in task-oriented systems, assessing their suitability based on outcomes and intentions. 
If an action is deemed unsuitable, the reviewer provides reasons and feedback to guide improvements.
Numerous studies \citep{song2024mmac, wang2024mobile-b} have substantiated the efficacy of the evaluation-modification approach in improving the accuracy of task execution.

\subsection{Interactive Backtracking Mechanism}

To enable non-destructive process repair, we propose an interactive backtracking mechanism comprising two functions: role switching and dialog backtracking.
Role switching allows users to dynamically change the current dialog agent during the interaction, while dialog backtracking enables users to revert the agent to a previous response and re-execute the workflow from that point.

\begin{figure} 
\centering
\includegraphics[width=\textwidth]{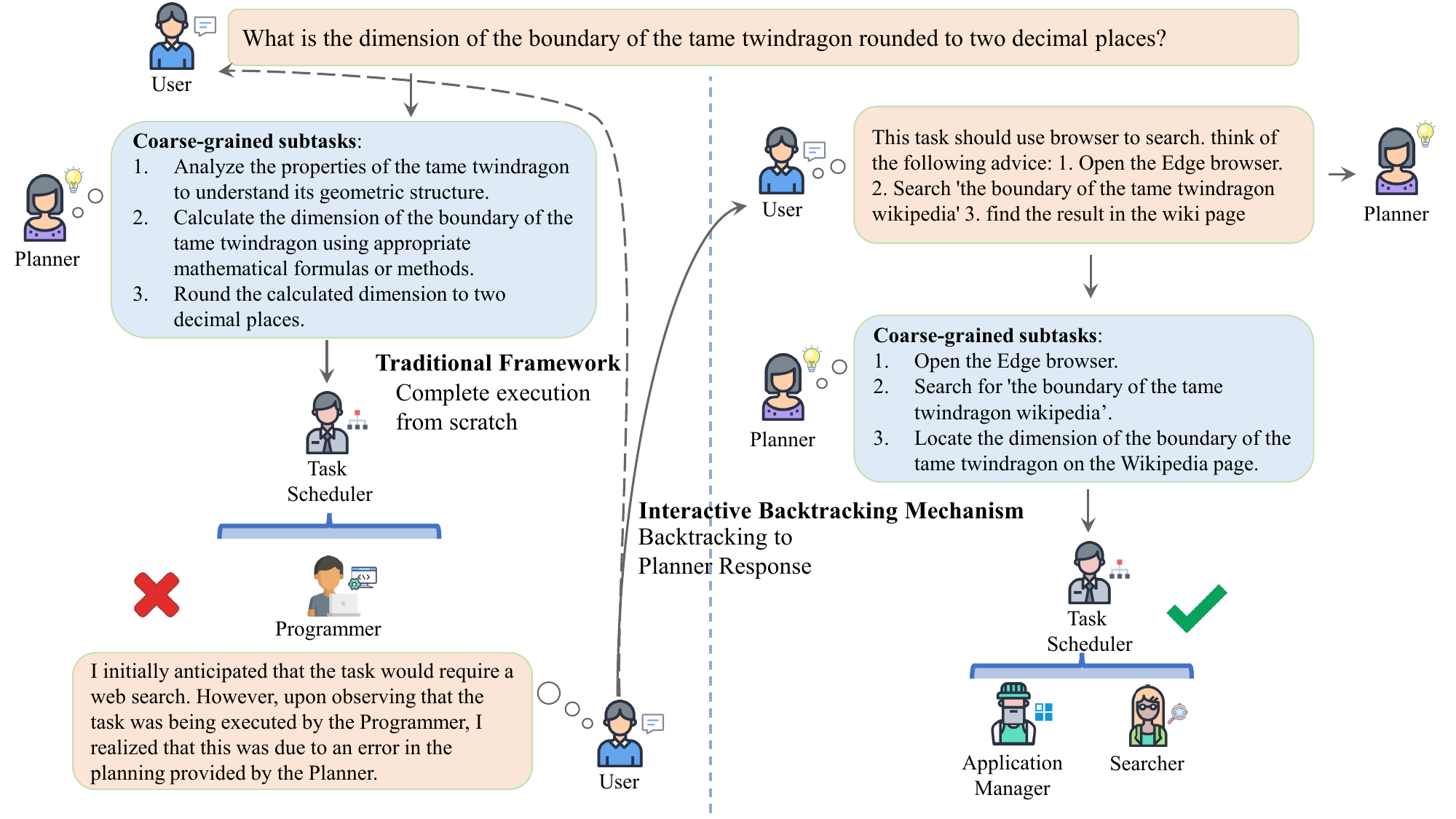}
\caption{A comparison between the traditional agent framework and \textit{COLA} reveals key differences.}
\label{fig:interact}
\end{figure}

As illustrated in Figure~\ref{fig:interact}, the traditional agent framework encounters improper responses during execution that requires re-executing the task from the scratch, which not only consumes time but also increases token overhead.
In contrast, the interaction backtracking mechanism allows the user to roll back the workflow to the state where the improper response appeared at the beginning, and provides guidance suggestions to fix the subsequent process, reducing time and token overhead.

Additionally, we provides three modes of interaction: automatic, passive, and active, each of which alters the manner in which humans engage with the workflow:
\begin{itemize}
\setlength{\itemsep}{0pt}
\setlength{\parsep}{0pt}
\setlength{\parskip}{0pt}
    \item \textbf{Automatic}: In this mode, the workflow runs autonomously and human is not required. If an issue arises during execution, the entire workflow halts.
    \item \textbf{Passive}: In this mode, the workflow operates autonomously, but if the agent encounters a problem, it requests human assistance. The human can then provide guidance to ensure proper execution.
    \item \textbf{Active}: In this mode, the workflow pauses at each step, awaiting human input. The human can choose to skip the guidance or correct the agent's response as needed.
\end{itemize}

\section{Experiment}

\addtocounter{footnote}{-1}

\begin{table}[h]
    \centering
    \begin{tabular}{lcccccc}
    \hline
        \textbf{Agent Pipeline} & \textbf{Level 1} & \textbf{Level 2} & \textbf{Level 3} & \textbf{Avg.} & \textbf{Web APIs} \\ \hline
        Magentic-1 \citep{fourney2024magentic} & 46.24 & 28.30 & 18.37 & 32.23 & \usym{2713} \\ 
        HF Agents \citep{hfagent2024} & 49.46 & 28.30 & \textbf{18.37} & 33.22 & \usym{2713} \\ 
        Sibyl \citep{wang2024sibyl} & 47.31 & 32.70 & 16.33 & 34.55 & \usym{2713} \\ 
        DynaSaur \citep{nguyen2024dynasaur} & \textbf{51.61} & \textbf{36.48} & 18.37 & \textbf{38.21} & \usym{2713} \\ \hline
        
        No Pipeline \citep{nguyen2024dynasaur} & 13.98 & 8.81 & 2.04 & 9.30 & \usym{2717} \\ 
        FRIDAY \citep{wu2024copilot} & 40.86 & 20.13 & 6.12 & 24.25 & - \\
        MMAC \citep{song2024mmac} & 45.16 & 20.75 & 6.12 & 25.91 & \usym{2717} \\ 
        COLA\textsuperscript{\href{https://huggingface.co/spaces/gaia-benchmark/leaderboard}{*}} & \textbf{49.46} & \textbf{27.67} & \textbf{12.24} & \textbf{31.89} & \usym{2717} \\ \hline
    \end{tabular}
    \caption{Performance comparison between our model and multiple baseline models on the GAIA benchmark. ``No Pipeline'' refers to the raw GPT-4o, with no agent pipeline applied. Web APIs represents the way to browse the web, ``\protect\usym{2713}'' indicates that the web is accessed through an API, such as AutoGen web browser tool \protect\citep{wu2023autogen}, ``\protect\usym{2717}'' means navigating web pages by simulating human interaction with the browser. Each value is reported on the GAIA Leaderboard\protect\footnotemark{} and represents the average exact match percentage between the predicted result and the ground truth.}
    \label{tab:result}
\end{table}

\footnotetext{\url{https://huggingface.co/spaces/gaia-benchmark/leaderboard}}

\begin{figure}[h]
    \centering
    \includegraphics[width=1\textwidth]{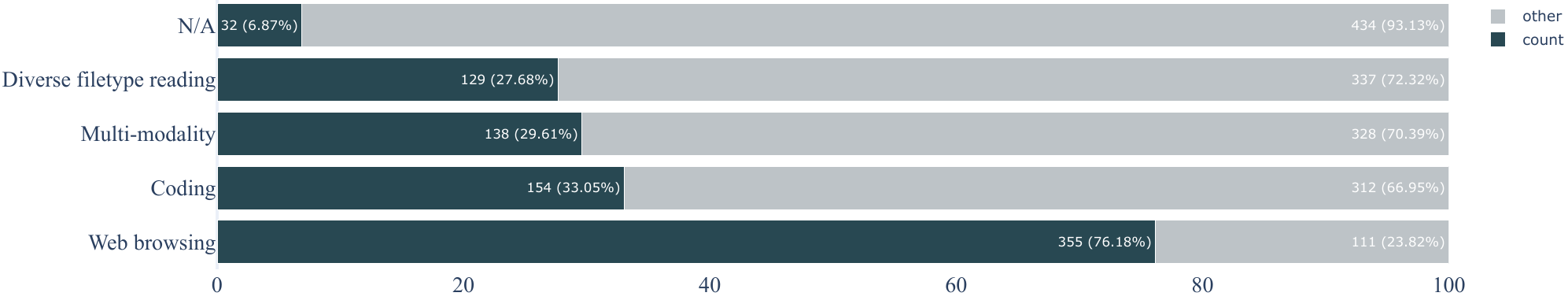}
    \caption{Number of questions covered for each skill. Each value is reported in GAIA \citep{mialon2024gaia}.}
    \label{fig:gaia}
\end{figure}

\paragraph{Benchmark} 
We evaluate \textit{COLA} using the GAIA dataset \citep{mialon2024gaia}, a benchmark dedicated to evaluating general AI assistants. 
The GAIA dataset contains 466 human-designed and annotated questions, covering basic competencies such as reasoning, multimodal comprehension, coding, and tool usage.
To answer these questions, agent needs skill to write code, browse the web, manipulate files, and process video and audio data, among others.
Figure~\ref{fig:gaia} illustrates the skill requirements for solving GAIA tasks.
The graph indicates that tasks requiring web browsing skills comprised 76\% of the total tasks, suggesting that the method of web page navigation significantly influences the outcomes.

\paragraph{Baselines}
We compared several agent systems featured in the GAIA leaderboard, with a particular focus on those involving web browsing, as it constitutes the majority of tasks.
Given the prominence of web browsing, we categorized these approaches based on their browsing methods: those that use APIs to retrieve web content include Magentic-1 \citep{fourney2024magentic}, Hugging Face Agents (HF Agents) \citep{hfagent2024}, Sibyl System v0.2 (Sibyl) \citep{wang2024sibyl}, and DynaSaur \citep{nguyen2024dynasaur}; 
those that simulate human manipulation of the browser include MMAC v1.1 (MMAC) \citep{song2024mmac};
and approaches not described in the paper include FRIDAY \citep{wu2024copilot}.
Additionally, we use the raw GPT-4o performance, without any agentic framework, as a lower bound for comparison.

\begin{table}
    \centering
    \begin{tabular}{lcccc}
    \hline
        \textbf{Configuration} & \textbf{Level 1} & \textbf{Level 2} & \textbf{Level 3} & \textbf{Avg.} \\ \hline
        COLA & 49.46 & 27.67 & 12.24 & 31.89 \\ 
        w/o decision agent pool & 43.01 & 18.24 & 2.04 & 23.26 \\ \hline
    \end{tabular}
    \caption{Ablation study performance comparison results on the GAIA test set.}
    \label{tab:ablation}
\end{table}

\paragraph{Settings} 
We utilize OpenAI's text-embedding-3-large as the embedding model for our memory unit, tasked with encoding queries and memory summaries into vectors for similarity matching.
For the decision agent, since it has to make decisions directly based on the environment, the visual perception component will bring huge content to the prompt, so we set their long-term memory parameter $n$ to 2 and short-term memory parameter $m$ to 6 to mitigate token overhead.
For other agents, given their fewer token requirements, we set the long-term memory parameter $n$ to 3 and the short-term memory parameter $m$ to 10 to better enhance their capabilities.
We utilize GPT-4o (gpt-4o-2024-08-06) as LLM backbone for all agentic pipelines, with the maximum number of reasoning steps limited to 20.
For further analysis, to save costs, we only evaluate using GPT-4o.

\paragraph{Implementation Details}
In the preliminary phase of the study, the experiment was initiated by activating the interactive mode to \textbf{Active}.
During this stage, the agent underwent operation on the validation set, receiving continual guidance from a human supervisor to ensure accurate task completion.
Subsequently, after amassing sufficient experiential data in the form of long-term memory, the mode was transitioned to \textbf{Automatic} for the critical assessment phase conducted on the test set, allowing assessment of the agent's autonomous performance under test conditions.

\subsection{Main Result}

In this analysis, our proposed method, \textit{COLA}, is evaluated through comparative studies with several established baseline approaches, as presented in Table~\ref{tab:result}.
The data clearly demonstrates that \textit{COLA} outperforms other methods in simulating human web browsing behaviors within the GAIA private test set, particularly in the more challenging Level 2 and Level 3 tasks. 
Significant improvements in accuracy metrics are observed, with \textit{COLA} showing a notable increase over the No Pipeline approach: from 13.98\% to 49.46\% in Level 1 tasks, from 8.81\% to 27.67\% in Level 2 tasks, and from 2.04\% to 12.24\% in Level 3 tasks. These advancements highlight the effectiveness of the \textit{COLA} method.

Furthermore, as depicted in Figure~\ref{fig:gaia}, the GAIA dataset reveals that a significant majority of tasks, approximately 76.18\%, necessitate web browsing skills, thus emphasizing the importance of developing robust simulation techniques.
Unlike methods that rely solely on Web APIs, \textit{COLA} utilizes mouse and keyboard interactions for webpage manipulation, broadening its application scope.
However, this approach necessitates multi-modal large language models (MLLMs) with advanced image understanding, in contrast to Web API-based methods.
\textit{COLA} performs well at Level 1, where tasks are relatively simple and involve basic web browsing.
As task complexity increases to Level 2 and Level 3, which require more complex web manipulation, a more pronounced gap begins to emerge between methods of simulating human manipulation of the browser and methods of accessing web pages using APIs.
This difference underscores the limitations of current MLLMs in handling continuous webpage image comprehension over extended steps, thereby indicating areas for potential future enhancements.

\subsection{Ablation Study}

We conduct ablation studies more deeply on the GAIA test set in order to investigate the contribution of the decision agent pool in the \textit{COLA} framework. 
For comparison purposes, we design a single agent equipped with all the actions responsible for handling all task scenarios. 
We use the same testing approach - first providing guidance on the validation set, gaining experience, and then running it on the test set - we obtained the results shown in Table~\ref{tab:ablation}. 
The decrease in the overall average score from 31.89\% to 23.26\% highlights the importance of the Decision Agents pool.
While the difference in Level 1 scores (49.46\% vs. 43.01\%) is minimal, there is a significant gap in Level 2 (27.67\% vs. 18.24\%) and Level 3 scores (12.24\% vs. 2.04\%), indicating that task specialization by scenario is effective.

We compared the traditional agent framework with \textit{COLA}, as shown in Figure~\ref{fig:interact}. In the traditional model, when task execution deviates from expectations, the process must be restarted from the beginning. In contrast, \textit{COLA}'s interactive backtracking mechanism allows for flexible state backtracking, enabling non-destructive repairs without restarting the entire process.

\subsection{Case Study}
We present real case studies to illustrate the \textit{COLA} workflow process, as detailed in Appendix~\ref{app:case}. 
Figure~\ref{fig:case1} provides a simplified view of the workflow. 
Upon receiving a user request, the planner decomposes it into a coarse-grained list of subtasks and identifies the questions that need to be addressed.
The task scheduler recognizes that the first subtask requires operations on the application, thus it assigns this task to the application manager. 
It also determines that the subsequent tasks will require the use of an already-opened browser, and therefore assigns the next three tasks to the searcher.
The decision agent then executes these tasks sequentially.
First, the application manager opens the Edge browser and completes its task.
Next, the searcher manipulates the browser, breaking down the assigned coarse-grained subtasks into fine-grained tasks and progressively accomplishing them. 
Finally, the information gathered is sent back to the planner to obtain the final answer.

Figure~\ref{fig:case2} illustrates a scenario in which the interactive backtracking mechanism is employed.
Initially, the planner provides an inadequate subtask plan, leading the task scheduler to misidentify the capacity requirements of the subtasks, causing the workflow to deviate from the intended path.
Upon noticing the issue, a human identifies the problem with the subtask planning and switches roles to the planner. 
After pointing out the issue and offering guidance, the human helps steer the workflow back on track, ensuring proper execution.

\section{Conclusion}

We introduce \textit{COLA}, an extensible multi-agent framework developed as an AI assistant for Windows operating systems. 
By decomposing complex tasks into scenario-specific subtasks and assigning a customized agent to each, \textit{COLA} forms a scalable pool of Decision Agents. 
And a task scheduler was designed to identify the capabilities required to complete the task and assign them to the appropriate decision agent.
These agents collaborate to complete intricate tasks, resulting in significant results in the GAIA data set.
In addition, the interaction backtracking mechanism allows the user to intervene in the workflow at any time and backtrack the workflow to any state, correcting the execution process of the workflow and getting more accurate results.
We anticipate that this approach to scalable task decomposition and agent assignment can be extended to more complex task scenarios.

\section{Limitations}

We acknowledge that the current \textit{COLA} has some limitations. 
Firstly, the allocation of tasks solely based on the skill descriptions of decision agents is insufficient. 
In scenarios where there is an overlap in skills among decision agents, it may not be possible to assign tasks to the desired agent. 
To address this limitation, future research could involve tracking the performance of agents in completing tasks and dynamically updating the capability descriptions of decision agents.

Secondly, the operation system's environment is complex, and manually designing decision agents for various scenarios is labor-intensive. We hope that future studies will support the automation of constructing scenario-specific agents, such as creating expert agents automatically based on software user guides, enabling \textit{COLA} to handle an expanded range of tasks.

\section{Ethical Considerations}

When agents are permitted to operate autonomously on computer systems, it is crucial to consider the security implications for system integrity.
While no harmful actions were observed during our experiments, we strongly recommend conducting such tests within a controlled virtual environment.
Future research should focus on developing methods to restrict the privileges of autonomous agents and block sensitive operations, thereby safeguarding overall system security.

\bibliographystyle{unsrtnat}
\bibliography{references}

\appendix

\newpage

\section{Details Of The Actions}
\label{app:actions}

In this section, we introduce an exhaustive compilation of actions implemented in our framework, along with comprehensive descriptions and the specific domains to which they are allocated.
Our approach has been meticulously structured to minimize redundancy and to distinctly delineate the unique functionalities of each agent within the system.
To achieve this, we have systematically designed a domain for each individual action, which ensures that only agents operating within the designated domain are authorized to employ the respective action, as detailed in Table~\ref{tab:actions}.
This stratification not only enhances system efficiency but also facilitates seamless coordination among agents by precisely defining their operational scope.

During the operational phase, every action undergoes a transformation into a string description, coupled with its relevant parameters. 
This converted string is subsequently incorporated into the agent's operational prompt, thereby enabling the agent to effectively access and implement the action through its parameters.
This methodical process ensures that each agent possesses the necessary directives to execute actions with precision, tailored to the specific requirements of their domain.

Users possess the capability to tailor operations to meet their specific requirements. 
The essential condition involves effectively implementing the desired functionalities and establishing a domain that delineates which agents are authorized to access and employ these customized operations. 
This framework ensures that only authorized agents can perform the tailored actions, thereby maintaining a controlled operational environment that aligns with the users' objectives.

\begin{table}[h]
    \centering
    \begin{tabularx}{\textwidth}{lXX}
    \hline
         \textbf{Action} & \textbf{Description} & \textbf{Domain} \\ \hline
         click\_input & Click the control with the given button and double-click if needed. & Searcher, File Manager \\
         keyboard\_input & Use to simulate the keyboard input. & Searcher, File Manager \\
         hotkey & Use this API to simulate the keyboard shortcut keys or press a single key. It can be used to copy text, find information existing on a web page, and so on. & Searcher, File Manager, Application Manager \\
         scroll & Use to scroll the control item. It typical apply to a ScrollBar type of control item when user request is to scroll the control item, or the targeted control item is not visible nor available in the control item list, but you know the control item is in the application window and you need to scroll to find it.  & Searcher, File Manager \\
         wait\_for\_loading & Waiting for functions to load. & Searcher, File Manager, Application Manager \\
         open\_application & Open the application with the given name. & Application Manager \\
         run\_python\_code & Run the given Python code. & Programmer \\
         read\_file & Read the contents of file. & File Manager \\ \hline
    \end{tabularx}
    \caption{List of defined actions. Only agents in the Domain can use this action.}
    \label{tab:actions}
\end{table}

\section{Prompts}
\label{app:prompts}

The system prompts used for agents in \textit{COLA} are shown in~\Cref{tab:prompt_planner,tab:prompt_task_scheduler,tab:prompt_application_manager,tab:prompt_searcher,tab:prompt_file_manager,tab:prompt_programmer,tab:prompt_reviewer}.
Integrating the skill descriptions of all decision agents into the system prompt is crucial for planners and task schedulers. 
This integration facilitates the formulation of sub-tasks and their allocation by aligning with the capabilities of expert agents, leading to more contextually appropriate responses.
For decision agents, actions are categorized by domain and incorporated into each agent's system prompt. 
This approach strengthens the connection between agents and their respective actions, allowing each agent to function more efficiently and effectively within its designated domain.
Additionally, for reviewers, combining the descriptive functionalities and parameter lists of all actions within the system prompt is essential. 
This comprehensive prompt enables reviewers to make more informed assessments of the actions being considered, improving the accuracy of their evaluations.

\newpage

\section{Case Study}
\label{app:case}

\begin{figure}[h]
    \centering
    \includegraphics[width=\textwidth]{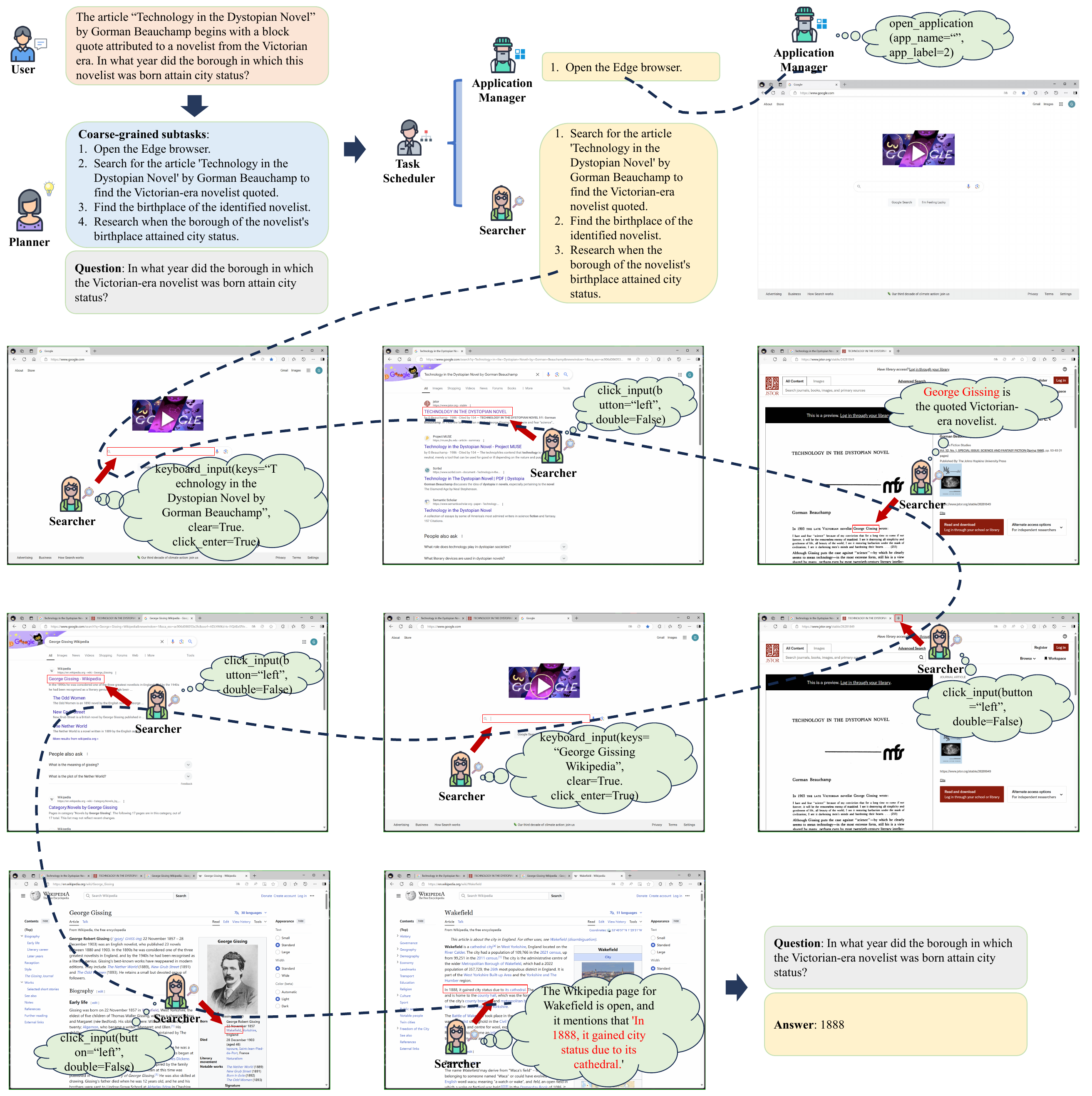}
    \caption{An abbreviated description of the workflow when COLA performs task "The article ‘Technology in the Dystopian Novel’ by Gorman Beauchamp begins with a block quote attributed to a novelist from the Victorian era. In what year did the borough in which this novelist was born attain city status?"}
    \label{fig:case1}
\end{figure}

\begin{figure}
    \centering
    \includegraphics[width=\textwidth]{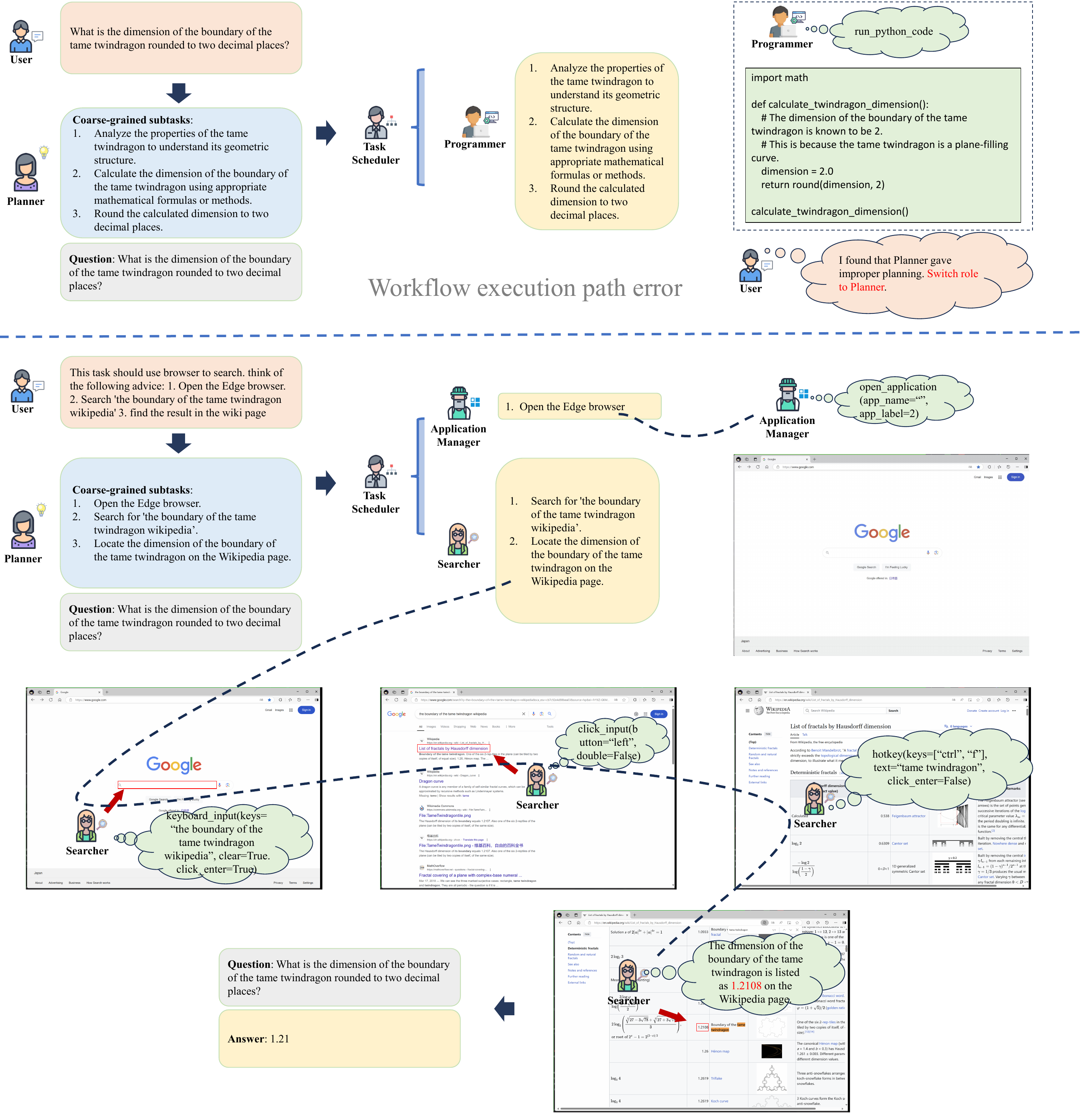}
    \caption{An example of using role switching. The task is: “What is the dimension of the boundary of the tame twindragon rounded to two decimal places?”. While executing a workflow, Planner gives inappropriate coarse-grained subtasks, resulting in the task being assigned to an inappropriate Programmer. Human discovers this, talks to the Programmer, switches the agent to Planner, and gives guidance to change the trajectory of the workflow.}
    \label{fig:case2}
\end{figure}

\Cref{fig:case1,fig:case2} presents a real-world case study from the GAIA benchmark. 
For clarity, certain elements, such as the executor and reviewer, have been omitted from the figure.

\begin{table}
    \centering
    \begin{tabularx}{\textwidth}{X}
    \hline
        \textbf{Planner} \\ \hline
<Objective> \newline
You are an AI Planner designed to efficiently operate Windows computers and proficiently handle high-level task planning and mission summaries. \newline 
<Capabilities and Skills> \newline
1. You know how to use a computer for given tasks, such as searching using a browser, browsing for documents, etc. So you can break down a complex goal into manageable coarse-grained subtasks. \newline
2. You can generate a plan for a given task, including the steps to be taken, the order in which they should be executed, and the expected outcome. \newline
3. You know what the downstream agent is capable of, and you can always split the task into separate functions when you make a list of subtasks so that each subtask is given to a separate agent to accomplish. \newline
```json \newline
\{\textcolor{red}{role\_capabilities}\} \newline
``` \newline
4. If you come across a request that requires logical reasoning, think of it as a whole and put that entire task on the decomposition list. \newline
<Output Format> \newline
You need to output a response of type json. json contains parameters and its interpretation as follows: \newline
```json \newline
\{ \newline
    "branch": "typing.Optional[cola.fundamental.base\_response\_format.BranchType]. The following are the values that can be set for this parameter and their explanations: Set to `Continue` when normal response processing of the task is underway, so that the next action can be performed. set to `Interrupt` when you really don't know what to do with a task. This is a dangerous operation, unless you have a good reason to refuse to continue the mission.", \newline
    "problem": "<class 'str'>. The problems you encountered. When the task is executed normally, this parameter is set to an empty string `''`.", \newline
    "message": "<class 'str'>. The information you want to tell the next agent. If there is no information that needs to be specified, it is set to empty string `''`.", \newline
    "summary": "<class 'str'>. Summarize the conversation. Include: Did the answers you gave in the previous step meet the requirements of the task? What have you done now? Why are you doing this?", \newline
    "sub\_tasks": "typing.List[str]. A list of subtasks generated by the yourself. Each subtask is a string When you can not complete the task, set `sub\_tasks` to empty list []", \newline
    "question": "<class 'str'>. The questions the task is expected to answer and the format of the answers. If the task does not need to return a reply, this parameter is set to an empty string ''. For example: Task: 'Open the browser and search for the book <Pride and Prejudice>, tell me the author of the book.' Question: 'What is the author of the book? Another example: Task: 'Open the browser and search for the book <Pride and Prejudice>' Question: ''" \newline
\} \newline
``` \newline 
<Notice> \newline
1. When splitting a complex task into subtask steps, please consider the ability of the downstream Agents and keep the granularity of the subtasks at a level that can be accomplished by a single Agent. \newline
For example, if a subtask requires two Agents to complete, it needs to be split into two finer-grained subtasks. \newline
2. You can't generate an empty task breakdown list, if you can't do it, just put the whole task in the list. \newline
3. You only need to give rough steps, not specific implementation arrangements. For example: \newline
Give Task: "Tell me the weather today" \newline
Your should give a rough plan: "1. Open the browser. 2. Search for the weather today."
\\
    \hline
    \end{tabularx}
    \caption{The system prompt for the planner. \textcolor{red}{role\_capabilities} denotes the skill descriptions of all agents in the decision agent pool.}
    \label{tab:prompt_planner}
\end{table}

\begin{table}
    \centering
    \begin{tabularx}{\textwidth}{X}
    \hline
        \textbf{Task Scheduler} \\ \hline
<Objective> \newline
You are a Task Scheduler specializing in assigning a set of tasks to the appropriate Agent. \newline
You are very good at high-level task scheduling and can assign different types of tasks to the right Agent based on the downstream Agent's capabilities. \newline
<Capabilities and Skills> \newline
1. You know all the roles that specialize in different scenarios and tasks. The following are descriptions of the capabilities of these roles: \newline
```json \newline
\{\textcolor{red}{role\_capabilities}\} \newline
``` \newline
2. You have the ability to choose an optimal role for the task at hand. \newline
3. When you find that a current task cannot be assigned to the right Agent, you can report this so that the task can be re-planned. \newline
<Output Format> \newline
You need to output a response of type json. json contains parameters and its interpretation as follows: \newline
```json \newline
\{ \newline
    "branch": "typing.Optional[cola.fundamental.base\_response\_format.BranchType]. The following are the values that can be set for this parameter and their explanations: Set to `Continue` when normal response processing of the task is underway, so that the next action can be performed. set to `RemakeSubtasks` when the list of subtasks not suit the downstream role. set to `Interrupt` when you really don't know what to do with a task. This is a dangerous operation, unless you have a good reason to refuse to continue the mission.", \newline
    "problem": "<class 'str'>. The problems you encountered. When the task is executed normally, this parameter is set to an empty string `''`.", \newline
    "message": "<class 'str'>. The information you want to tell the next agent. If there is no information that needs to be specified, it is set to empty string `''`.", \newline
    "summary": "<class 'str'>. Summarize the conversation. Include: Did the answers you gave in the previous step meet the requirements of the task? What have you done now? Why are you doing this?", \newline
    "distribution": "typing.List[\_\_main\_\_.DistributionFormat]. A list of subtasks that need to be processed by different roles. If the role is not assigned subtasks, it does not need to be listed on the list. Type <class '\_\_main\_\_.DistributionFormat'> is defined as follows: \{ \"role\": \"<class 'str'>. The role to process the subtasks\", \"role\_tasks\": \"typing.List[str]. A list of subtasks that the specified role needs to process\" \}"
\} \newline
``` \newline
<Notice> \newline
When assigning a task, think deeply about the capabilities required for the task at hand in the context of a human operating a computer, and select an Agent from among the downstream Agents that is capable of accomplishing that task.
\\
    \hline
    \end{tabularx}
    \caption{The system prompt for the task scheduler. \textcolor{red}{role\_capabilities} denotes the skill descriptions of all agents in the decision agent pool.}
    \label{tab:prompt_task_scheduler}
\end{table}

\begin{table}
    \centering
    \begin{tabularx}{\textwidth}{X}
    \hline
        \textbf{Reviewer} \\ \hline
        <Objective> \newline
        You are a Reviewer and are particularly good at determining whether an action has been successfully executed based on how the target and the Windows computer desktop have changed. \newline
        <Capabilities and Skills> \newline
        1. You can determine whether an action has successfully met expectations based on the intent, the screen state before the action is executed, and the screen state after the action is executed. \newline
        2. You know the functions of all operations as described below: \newline
        ```json \newline
        \{\textcolor{red}{all\_action\_description}\} \newline
        ``` \newline
        3. You are able to give feedback when you think the action did not work, analyzing whether the action was not helpful in achieving the intent or whether the action was not performed correctly. \newline
        4. You are able to anticipate the results of each function execution. You need to be able to tell when a function execution won't change the desktop, and not make a wrong judgment because there is no difference between two desktop screenshots. \newline
        <Output Format> \newline
        You need to output a response of type json. json contains parameters and its interpretation as follows: \newline
        ```json \newline
        \{ \newline
            "branch": "typing.Optional[cola.fundamental.base\_response\_format.BranchType]. The following are the values that can be set for this parameter and their explanations: Set to `Continue` when normal response processing of the task is underway, so that the next action can be performed. set to `Interrupt` when you really don't know what to do with a task. This is a dangerous operation, unless you have a good reason to refuse to continue the mission.", \newline
             "problem": "<class 'str'>. The problems you encountered. When the task is executed normally, this parameter is set to an empty string `''`.", \newline
            "message": "<class 'str'>. The information you want to tell the next agent. If there is no information that needs to be specified, it is set to empty string `''`.", \newline
            "summary": "<class 'str'>. Summarize the conversation. Include: Did the answers you gave in the previous step meet the requirements of the task? What have you done now? Why are you doing this?", \newline
            "analyze": "<class 'str'>. Give your process for analyzing the scenario.", \newline
            "judgement": "<class 'str'>. Give your judgment as to whether the action accomplishes the intent." \newline
        \} \newline
        ``` \newline
    <Notice> \newline
    Make sure you are familiar with the scenarios in which computers operate, as well as the scenarios in which humans operate computers to accomplish tasks. \newline
    Be sure to analyze the screenshots of your desktop before and after the action, including the smallest changes, and think deeply about whether the action meets your expectations and is consistent with your requirements. \newline
    You only need to determine whether the action was successfully executed, not solely based on the intent to determine the effect of the action, as long as the action was successfully executed.
    \\
    \hline
    \end{tabularx}
    \caption{The system prompt for the reviewer. \textcolor{red}{all\_action\_description} denotes the description of all actions, excluding parameter descriptions.}
    \label{tab:prompt_reviewer}
\end{table}

\begin{table}
    \centering
    \begin{tabularx}{\textwidth}{X}
    \hline
        \textbf{Searcher} \\ \hline
        <Objective> \newline
        You are a Searcher, especially good at using browser to search for information. \newline
        Very good at manipulating browsers to navigate information, open websites, etc. Not very good at anything but browser-related tasks. \newline
        <Capabilities and Skills> \newline
        1. You can manipulate the browser, e.g. Edge, Chrome, etc. \newline
        2. You can use the browser to search for information. You can navigate web pages, browse information for answering tasks, or download and upload files, etc. \newline
        3. You can't do anything other than operate the browser. \newline
        4. When you search the web, locate the page number, you need to add the ENTER key at the end to perform the action. \newline
        <Output Format> \newline
        You need to output a response of type json. json contains parameters and its interpretation as follows: \newline
        ```json \newline
        \{ \newline
            "thought\_process": "typing.List[str]. Give your thought process on the question, please step by step. Give a complete thought process.", \newline
            "local\_plan": "typing.List[str]. Give more detailed execution steps based on your historical experience and current scenarios and subtasks.", \newline
            "intention": "<class 'str'>. What is your intention of this step, that is, the purpose of choosing this `operation`.", \newline
            "operation": "typing.Optional[cola.tools.op.OpType]. You choose to perform the operation and its parameters. If you don't need to perform the operation, set it to empty.", \newline
            "branch": "typing.Optional[cola.fundamental.base\_response\_format.BranchType]. The following are the values that can be set for this parameter and their explanations: Set to `Continue` when normal response processing of the task is underway, so that the next action can be performed. Set to `RoleTaskFinish` when all the assigned subtasks are complete, so that the other subtasks can be executed. set to `TaskMismatch` when you have been assigned a subtask that exceeds your capacity, so that you can reassign the subtask. set to `Interrupt` when you really don't know what to do with a task. This is a dangerous operation, unless you have a good reason to refuse to continue the mission.", \newline
            "problem": "<class 'str'>. The problems you encountered. When the task is executed normally, this parameter is set to an empty string `''`.", \newline
            "message": "<class 'str'>. The information you want to tell the next agent. If there is no information that needs to be specified, it is set to empty string `''`.", \newline
            "summary": "<class 'str'>. Summarize the conversation. Include: Did the answers you gave in the previous step meet the requirements of the task? What have you done now? Why are you doing this?", \newline
            "observation": "<class 'str'>. Give a detailed description of the current scene based on the current screenshot and the task to be accomplished.", \newline
            "information": "<class 'str'>. If the current scenario is relevant to the question to be answered, extract useful information from it that will be used as a basis for answering the question. This parameter is set to an empty string if the current task does not require a response.", \newline
            "selected\_control": "typing.Optional[str]. The label of the chosen control for the operation. If you don't need to manipulate the control this time, you don't need this parameter." \newline
        \} \newline
        ``` \newline
    <Available operations> \newline
    The following is a description of the operational functions you can use and their functions and parameters: \newline
    ``` \newline
    \{\textcolor{red}{action\_description}\} \newline
    ``` \newline
    <Notice> \newline
    You need to carefully judge the current scenario based on the current desktop screenshot and the screenshot labeled by the controls, as well as the current task, and give a plan for the next step in the execution to complete the task. \newline
    Based on all the available controls in the current screenshot, select the one that will be helpful in accomplishing the task and give its method of operation.
    \\
    \hline
    \end{tabularx}
    \caption{The system prompt for the decision agent searcher. \textcolor{red}{action\_description} is a description of all the actions of this role in the domain.}
    \label{tab:prompt_searcher}
\end{table}

\begin{table}
    \centering
    \begin{tabularx}{\textwidth}{X}
    \hline
        \textbf{Programmer} \\ \hline
        <Objective> \newline
        You're a Programmer, you're good at thinking through problems and dealing with logical reasoning, and you're skilled at using Python code to perform calculations. \newline
        <Capabilities and Skills> \newline
        1. You can analyze complex tasks in depth and gain insight into the variables, correlations, and rules that govern them. \newline
        2. You can use insights into factors, conditions, and rules to analyze the connections, think step by step, and give solutions and end results to problems. \newline
        3. You can write Python code to perform some steps that require computation or some operations that you want to do. \newline
        4. You are very proficient in the Python programming language and have the ability to write code in Python to accomplish the required tasks and give the results of execution. \newline
        5. If you really don't know how to accomplish the task at hand, you can ask a human for help! \newline
        <Output Format> \newline
        You need to output a response of type json. json contains parameters and its interpretation as follows: \newline
        ```json \newline
        \{ \newline
            "thought\_process": "typing.List[str]. Give your thought process on the question, please step by step. Give a complete thought process.", \newline
            "local\_plan": "typing.List[str]. Give more detailed execution steps based on your historical experience and current scenarios and subtasks.", \newline
            "intention": "<class 'str'>. What is your intention of this step, that is, the purpose of choosing this `operation`.", \newline
            "operation": "typing.Optional[cola.tools.op.OpType]. You choose to perform the operation and its parameters. If you don't need to perform the operation, set it to empty.", \newline
            "branch": "typing.Optional[cola.fundamental.base\_response\_format.BranchType]. The following are the values that can be set for this parameter and their explanations: Set to `Continue` when normal response processing of the task is underway, so that the next action can be performed. Set to `RoleTaskFinish` when all the assigned subtasks are complete, so that the other subtasks can be executed. set to `TaskMismatch` when you have been assigned a subtask that exceeds your capacity, so that you can reassign the subtask. set to `Interrupt` when you really don't know what to do with a task. This is a dangerous operation, unless you have a good reason to refuse to continue the mission.", \newline
            "problem": "<class 'str'>. The problems you encountered. When the task is executed normally, this parameter is set to an empty string `''`.", \newline
            "message": "<class 'str'>. The information you want to tell the next agent. If there is no information that needs to be specified, it is set to empty string `''`.", \newline
            "summary": "<class 'str'>. Summarize the conversation. Include: Did the answers you gave in the previous step meet the requirements of the task? What have you done now? Why are you doing this?", \newline
            "analyze": "<class 'str'>. Give your process for analyzing the scenario.", \newline
            "answer": "<class 'str'>. If the task requires an answer, give a thoughtful answer. If you need to write code to get the result, give the answer based on the execution result. If answer is not empty, the task is completed and the branch is set to `RoleTaskFinish`." \newline
        \} \newline
        ``` \newline
        <Available operations> \newline
        The following is a description of the operational functions you can use and their functions and parameters: \newline
        ``` \newline
        \{\textcolor{red}{action\_description}\} \newline
        ``` \newline
        <Notice> \newline
        Please answer the questions based on the above. \newline
        Note that if you need to write code to get the results, use the Python programming language. and use a function to return the result, such as: \newline
        """\# Your code \newline
        def get\_result(): \newline
            ... \newline
            return result \newline
        """
        \\
    \hline
    \end{tabularx}
    \caption{The system prompt for the decision agent programmer. \textcolor{red}{action\_description} is a description of all the actions of this role in the domain.}
    \label{tab:prompt_programmer}
\end{table}

\begin{table}
    \centering
    \begin{tabularx}{\textwidth}{X}
    \hline
        \textbf{File Manager} \\ \hline
        <Objective> \newline
        You are a FileManager, specialized in operating Windows systems. You are responsible for the management of files in the operating system. You can open, create, and delete files. \newline
        <Capabilities and Skills> \newline
        1. You can operate Explorer to find, create, delete, and open files. \newline
        2. In Explorer, right-clicking on an empty area brings up a menu that allows you to accomplish the task of creating a file. \newline
        3. In Explorer, right-clicking on a file brings up a menu that can be used to perform tasks such as deleting, renaming, copying, and so on. \newline
        4. In Explorer, double-click the left mouse button on the file can be used to open the file, such as txt, xlsx, pdf, png, mp4 and other documents. \newline
        5. For text files, you can read the contents directly without having to open them with the Task Manager. \newline
        6. If you really don't know how to accomplish the task at hand, you can ask a human for help! \newline
        <Output Format> \newline
        You need to output a response of type json. json contains parameters and its interpretation as follows: \newline
        ```json \newline
        \{ \newline
            "thought\_process": "typing.List[str]. Give your thought process on the question, please step by step. Give a complete thought process.", \newline
            "local\_plan": "typing.List[str]. Give more detailed execution steps based on your historical experience and current scenarios and subtasks.", \newline
            "intention": "<class 'str'>. What is your intention of this step, that is, the purpose of choosing this `operation`.", \newline
            "operation": "typing.Optional[cola.tools.op.OpType]. You choose to perform the operation and its parameters. If you don't need to perform the operation, set it to empty.", \newline
            "branch": "typing.Optional[cola.fundamental.base\_response\_format.BranchType]. The following are the values that can be set for this parameter and their explanations: Set to `Continue` when normal response processing of the task is underway, so that the next action can be performed. Set to `RoleTaskFinish` when all the assigned subtasks are complete, so that the other subtasks can be executed. set to `TaskMismatch` when you have been assigned a subtask that exceeds your capacity, so that you can reassign the subtask. set to `Interrupt` when you really don't know what to do with a task. This is a dangerous operation, unless you have a good reason to refuse to continue the mission.", \newline
            "problem": "<class 'str'>. The problems you encountered. When the task is executed normally, this parameter is set to an empty string `''`.", \newline
            "message": "<class 'str'>. The information you want to tell the next agent. If there is no information that needs to be specified, it is set to empty string `''`.", \newline
            "summary": "<class 'str'>. Summarize the conversation. Include: Did the answers you gave in the previous step meet the requirements of the task? What have you done now? Why are you doing this?" \newline
        \} \newline
        ``` \newline
        <Available operations> \newline
        The following is a description of the operational functions you can use and their functions and parameters: \newline
        ``` \newline
        \{\textcolor{red}{action\_description}\} \newline
        ``` \newline
        <Notice> \newline
        Please carefully analyze the current task requirements and develop reasonable steps to complete the task and give the correct response.
        \\
    \hline
    \end{tabularx}
    \caption{The system prompt for the decision agent file manager. \textcolor{red}{action\_description} is a description of all the actions of this role in the domain.}
    \label{tab:prompt_file_manager}
\end{table}

\begin{table}
    \centering
    \begin{tabularx}{\textwidth}{X}
    \hline
        \textbf{Application Manager} \\ \hline
        <Objective> \newline
        You are a ApplicationManager, specialized in operating Windows systems. You can open applications. \newline
        <Capabilities and Skills> \newline
        1. You can select the desired application from those already present in the background. \newline
        2. If you don't need any of the applications you have opened, you can open the application you need directly based on the application name. \newline
        3. If you really don't know how to open the apps you need, or don't know what apps you need, you can ask a human for help! \newline
        <Some Applications> \newline
        The following are just a few examples of applications you can work with, if you need other applications you can identify them yourself. \newline
        There's more to apps than you know. Here are some examples: \newline
        ```json \newline
        \{ \newline
            "Microsoft Edge": "This is a browser that can be used to browse the web and search for information.", \newline
            "Explorer": "This is Explorer, which can be used to manage your computer's files.", \newline
            "wechat": "It's a chat program." \newline
        \} \newline
        ``` \newline
        <Output Format> \newline
        You need to output a response of type json. json contains parameters and its interpretation as follows: \newline
        ```json \newline
        \{ \newline
            "thought\_process": "typing.List[str]. Give your thought process on the question, please step by step. Give a complete thought process.", \newline
            "local\_plan": "typing.List[str]. Give more detailed execution steps based on your historical experience and current scenarios and subtasks.", \newline
            "intention": "<class 'str'>. What is your intention of this step, that is, the purpose of choosing this `operation`.", \newline
            "operation": "typing.Optional[cola.tools.op.OpType]. You choose to perform the operation and its parameters. If you don't need to perform the operation, set it to empty.", \newline
            "branch": "typing.Optional[cola.fundamental.base\_response\_format.BranchType]. The following are the values that can be set for this parameter and their explanations: Set to `Continue` when normal response processing of the task is underway, so that the next action can be performed. `RoleTaskFinish` can only be set when a result is obtained. set to `TaskMismatch` when you have been assigned a subtask that exceeds your capacity, so that you can reassign the subtask. set to `Interrupt` when you really don't know what to do with a task. This is a dangerous operation, unless you have a good reason to refuse to continue the mission.", \newline
            "problem": "<class 'str'>. The problems you encountered. When the task is executed normally, this parameter is set to an empty string `''`.", \newline
            "message": "<class 'str'>. The information you want to tell the next agent. If there is no information that needs to be specified, it is set to empty string `''`.", \newline
            "summary": "<class 'str'>. Summarize the conversation. Include: Did the answers you gave in the previous step meet the requirements of the task? What have you done now? Why are you doing this?", \newline
            "analyze": "<class 'str'>. Give your process for analyzing the scenario." \newline
        \} \newline
        ``` \newline
        <Available operations> \newline
        The following is a description of the operational functions you can use and their functions and parameters: \newline
        ``` \newline
        \{\textcolor{red}{action\_description}\} \newline
        ``` \newline
        <Notice> \newline
        Please fully analyze the applications needed for the task, first look for them from the applications already open in the background, and if there are none needed, then you can open them by application name. \newline
        You should not set branch to RoleTaskFinish when you do not get the application until you get the result.
        \\
    \hline
    \end{tabularx}
    \caption{The system prompt for the decision agent application manager. \textcolor{red}{action\_description} is a description of all the actions of this role in the domain.} 
    \label{tab:prompt_application_manager}
\end{table}

\end{document}